# KTlO: A metal shrouded 2D semiconductor with high carrier mobility and tunable magnetism

Ya-Qian Song,[1,+] Jun-Hui Yuan,[1,+] Li-Heng Li,[1] Ming Xu,[1] Jia-Fu Wang,[2] Kan-Hao Xue,[1,3*] and Xiang-Shui Miao[1]

[1] Wuhan National Research Center for Optoelectronics, School of Optical and Electronic Information, Huazhong University of Science and Technology, Wuhan 430074, China

[2] School of Science, Wuhan University of Technology, Wuhan 430070, China

[3] IMEP-LAHC, Grenoble INP – Minatec, 3 Parvis Louis Néel, 38016 Grenoble Cedex 1, France

[+]The authors Y.-Q. Song and J.-H. Yuan contributed equally to this work.

## Corresponding Author

*E-mail: xkh@hust.edu.cn (K.-H. Xue)

## ABSTRACT

Two-dimensional (2D) materials with high carrier mobility and tunable magnetism are in high demand for nanoelectronics and spintronic applications. Herein, we predict a novel two-dimensional monolayer KTlO that possesses an indirect band gap of 2.25 eV (based on HSE06) and high carrier mobility ($1.86 \times 10^3$ cm$^2$ V$^{-1}$s$^{-1}$ for electron and $2.54 \times 10^3$ cm$^2$ V$^{-1}$s$^{-1}$ for hole) by means of *ab initio* calculations. KTlO monolayer has a calculated cleavage energy of 0.56 J m$^{-2}$, which suggests exfoliation of bulk material as viable means for the preparation of mono- and few-layer materials. Remarkably, the KTlO monolayer suggests tunable magnetism and half-metallicity with hole doping, which are attributed to the novel Mexican-hat-like bands and van Hove singularities in its electron structure. Furthermore, monolayer KTlO exhibits moderate optical absorption over visible light and ultraviolet region. The band gap value and band characteristics of monolayer KTlO can be strongly manipulated by biaxial and uniaxial strains to meet the requirements of various applications. All these novel properties render monolayer KTlO a promising functional material for future nanoelectronics and spintronic applications.

**KEYWORDS:** *2D materials, monolayer KTlO, mobility, magnetism, electronic properties, hole-doping, density functional theory*

## I. Introduction

Two-dimensional (2D) materials have attracted enormous attention since the successful mechanical exfoliation of graphene in 2004.[1–3] To date, the family of 2D materials is growing rapidly, including elemental monolayers (such as group-III, group-IV, group-V),[4–11] MXenes,[12–14] transitional metal dichalcogenides (TMDCs),[15–18] metal oxides[19–22] and so forth.[23–25] These 2D materials exhibit extraordinary properties that have been studied in various fields, such as field-effect transistors, photovoltaic solar cells and optoelectronic devices.[16,26,27] In particular, 2D materials with local magnetic moments hold great potential for spintronic applications, such as spin field effect transistors, spin light-emitting diodes and solid-state quantum information processing devices.[28–32] Besides using intrinsically magnetic materials, there are thus far several approaches to induce magnetism, such as introducing adatoms, defects and edges to the system. However, these approaches face serious challenges in experiments, affected by factors such as structural disorder, uncontrollable concentration of dopants and vacancies. On the other hand, system with a Mexican-hat-like valence band maximum (VBM) may give rise to high density of states (DOS) and an almost one-dimensional-like van Hove singularity near the VBM. In these systems, hole-doping may induce a spontaneous ferromagnetic transition, observed in GaSe, $\alpha$-SnO and InP$_3$ monolayers.[33–35] Compared with conventional approaches such as transition metal element doping, the magnetism in these 2D materials with Mexican-hat-like bands is inherent and can be tuned by the doping level, through liquid electrolyte gating.[36,37] To this end, searching for new candidate 2D materials with high carrier mobility and tunable magnetism is of great interest in spintronic devices.

In this work, using first-principles calculations we report a new monolayer metal shrouded semiconductor KTlO, with high dynamic and thermal stability. In addition, KTlO shows remarkably weak interlayer interactions, which result in a relatively low cleavage energy of 0.56 J m$^{-2}$. With an indirect band gap of 2.25 eV, monolayer KTlO shows high carrier mobilities of $2.54 \times 10^3$ cm$^2$ V$^{-1}$ s$^{-1}$ for electrons and $1.86 \times 10^3$ cm$^2$ V$^{-1}$ s$^{-1}$ for holes. Most fascinatingly, it exhibits extended singularity points in the DOS near the Mexican-hat shaped VBM, as well as the half-metallicity that can be tuned by hole doping. Such gate-controlled magnetism and high mobilities for electron and hole carriers in 2D KTlO render it a particularly strong candidate for spintronic devices.

## II. Computational methods

All DFT calculations were performed using the plane-wave-based Vienna *Ab-initio* Simulation Package (VASP) code.[38,39] The generalized gradient approximation (GGA) within the Perdew-Burke-Ernzerhof (PBE)[40] functional form was used for the exchange-correlation energy, and projector augmented-wave (PAW) pseudopotentials[41,42] were used to replace the core electrons. The Heyd–Scuseria–Ernzerhof (HSE06) screened hybrid functional[35] has been used to calculate the band structures of KTlO in order to rectify the underestimated band gaps in PBE. The plane wave energy cutoff was fixed to be 500 eV. Description of the van der Waals (vdW) interactions was corrected using the DFT-D3 approach.[43] For all structural relaxations, the convergence criterion for total energy was set to $1.0 \times 10^{-6}$ eV, and structural optimization was obtained until the Hellmann-Feynman force acting on each atom was less than 0.01 eV/Å for any direction. The phonon dispersion relations were calculated with the density functional perturbation theory, using the PHONOPY code.[44] *Ab initio* molecular dynamics (AIMD) simulations were performed to check the thermal stability of the structures, where the NVT canonical ensemble was used.

## III. Results and discussions

As shown in **Fig. 1(a)**, the symmetry for bulk KTlO was found to be monoclinic with the space group $C2/m$, and two KTlO layers constitute the unit cell. The optimized lattice parameters of bulk KTlO were $a = 13.10$ Å, $b = 3.66$ Å, $c = 6.44$ Å and $\beta = 108.15°$, in good accordance with the reported experimental results ($a = 12.87$ Å, $b = 3.62$ Å and $c = 6.29$ Å, $\beta = 106.59°$).[45] The two KTlO layers were joined together by vdW interactions with a interlayer distance of 3.59 Å. The layered KTlO shows a typical buckling structure, similar to $Tl_2O$ and $K_2O$,[46,47] while the buckling height is 3.13 Å. Interestingly, KTlO may be regarded as a 'mixture' of $Tl_2O$ and $K_2O$ with an alternately ordered combination. In addition, bulk KTlO exhibits an indirect band gap feature, where the value was calculated to be 1.46 eV and 2.01 eV at the PBE and HSE06 levels (**Fig. 1(b)**), respectively. The HSE06 gap value can be regarded as a good approximation to the true fundamental gap.

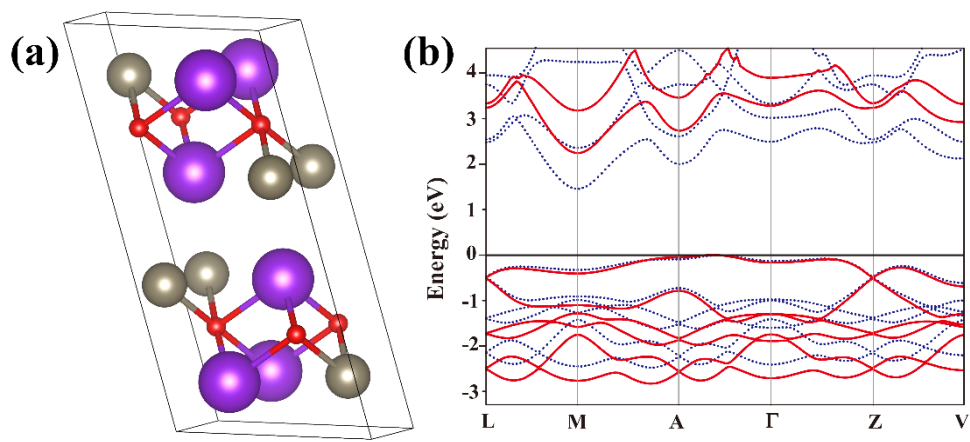

**Figure 1.** (a) Crystal structure of optimized bulk KTlO, where big purple balls, medium brown balls and small red balls represent K, Tl and O atoms, respectively. (b) Calculated electronic band structures of bulk KTlO using PBE (dash line) and HSE06 (solid line) functionals, respectively. The Fermi level is set to zero energy.

Monolayer KTlO has been obtained in our simulation by taking an atomic layer from the KTlO bulk along the [001] direction, as shown in **Fig. 2(a)**. It exhibits a rectangular configuration with optimized lattice parameters of $a = 3.80$ Å and $b = 6.30$ Å. The K-O bond lengths are 2.67 Å and 2.71 Å in monolayer KTlO, while the Tl-O bond lengths are 2.54 Å and 2.40 Å, being shorter than that of $Tl_2O$ (2.57 Å).[46] The buckling height of monolayer KTlO is 2.98 Å, slightly smaller than that of bulk. Indeed, the crystal structure of the 'mixture' KTlO demonstrates substantial deviations comparing with the 'clean' $Tl_2O$ and $K_2O$.

As mentioned above, KTlO shows a typical van der Waals stacking structure, thus mechanical or liquid phase exfoliation may be possible just as for graphene and black phosphorus[1,10,48]. To assess such possibility, we calculated the cleavage energy of monolayer KTlO from a five-layer KTlO slab, serving as a model of the bulk. As shown in **Fig. 2(b)**, the cleavage energy increases with the interlayer distance, reaching a converged value of 0.56 J m$^{-2}$. Our estimated exfoliation energies of graphene and black phosphorus are 0.32 J m$^{-2}$ and 0.37 J m$^{-2}$, respectively, consistent with previous theory studies.[33,49] Therefore, it is feasible to obtain monolayer KTlO through exfoliation from the bulk, as the cleavage energy is in the same range of common 2D materials.

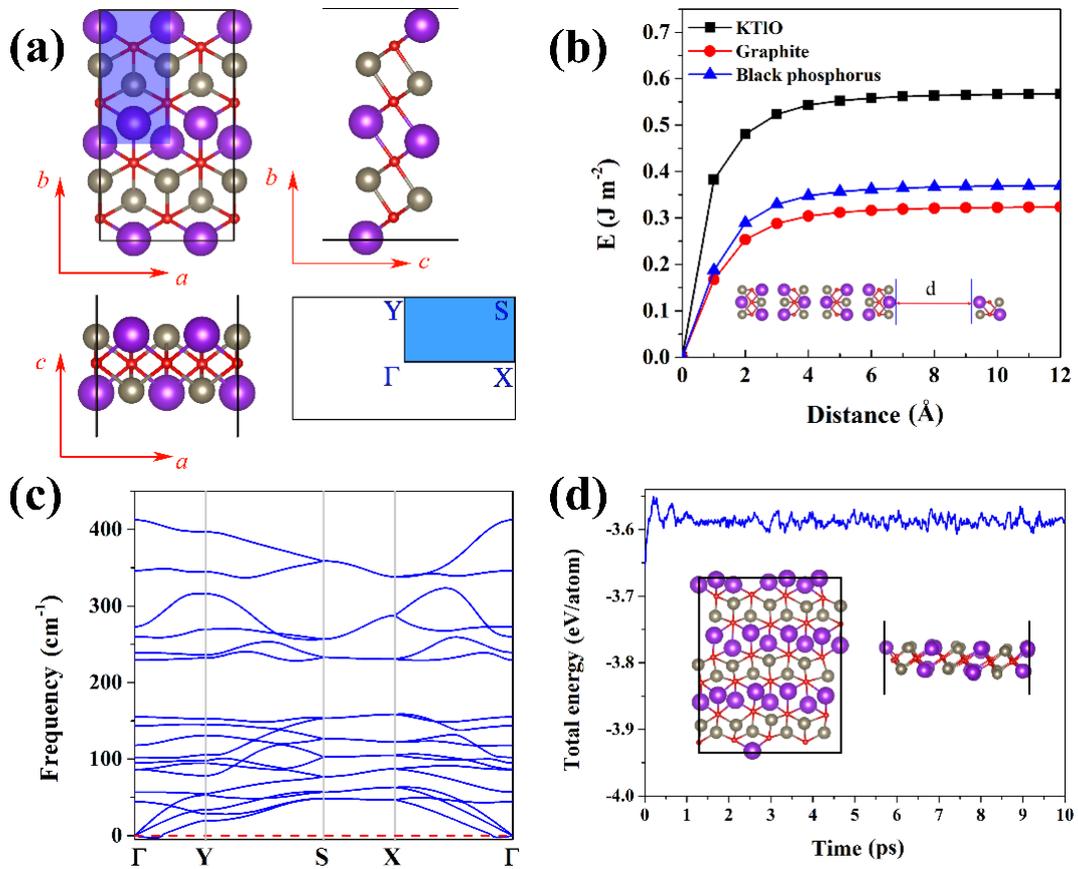

**Figure 2.** (a) Top and side views of a 2×2 monolayer KTlO supercell, and the corresponding first Brillouin zone with high symmetry $k$ points. The light blue square marks a unit cell of KTlO. (b) Cleavage energy estimation for the formation of monolayer KTlO, calculated by gradually enlarging the interlayer distance between a selected surface monolayer and the remainder of a five-layer slab, resembling the exfoliation process of a monolayer from the bulk model. (c) The calculated phonon dispersion spectra of monolayer KTlO. (d) Side view of the structure snapshots from the molecules dynamic simulation of the KTlO monolayer at 500 K, as well as the variation of total energy over the 10 *ps* simulation time.

In addition, the phonon dispersions of monolayer KTlO (shown in **Fig. 2(c)**) only show a tiny imaginary phonon mode (2.8 cm$^{-1}$) near the $\Gamma$ point, which comes from systematic computational error, indicating the kinetic stability. The thermal stability is further substantiated by AIMD simulations (see **Fig. 2(d)**), where the monolayer KTlO structure remains intact at 500 K after a 10 *ps* simulation time.

After verifying the stability and the feasibility of exfoliation, we turn our attention to the electronic properties of monolayer KTlO. First, to understand the bonding characteristics, we calculated its electron localization function (ELF)[50–53] and Bader charges.[54–56] As shown in **Fig. 3(a)**, ELF = 1 corresponds to perfect localization, ELF = 0.5 corresponds to the free electron gas and ELF = 0 means the absence of electrons. For monolayer KTlO, the electron is localized around the atoms, and nearly zero along the Tl-O and K-O bonds, indicating typical ionic bond nature of Tl-O and K-O. On the other hand, the Bader charge analysis shows that 0.807 e has been transferred from a K atom to its surrounding O atoms, while the amount of charge transferred from a Tl atom is 1.01 e. For the element Tl, it can feature two stable oxidation states Tl$^+$ and Tl$^{3+}$, such as Tl$_2$O and Tl$_2$O$_3$. Yet, in monolayer KTlO, Tl prefers to be stabilized mainly as Tl$^+$ cations, similar to that of Tl$_2$O.[46] In addition, there is also extra valence electrons held in Tl atoms, which results in high electron localization around Tl atoms, as shown in **Fig. 3(a)**. The ionic bonding nature in KTlO also implies that the effect of spin-orbit coupling (SOC) should be minor in the KTlO system, since as the only heavy element, the Tl cation has lost most of its 6*p* electrons. To verify this argument, we examined the band structures of monolayer KTlO with and without SOC using PBE and HSE06 functionals (Fig. 3(b) and Fig. S1). According to our calculations, the difference in band gap value is less than 0.02 eV upon switching on SOC. On the other hand, the VBM and conduction band minimum (CBM) characteristics remain the same, either with or without considering SOC. Therefore, we shall neglect the SOC effect in all forthcoming band structure calculations.

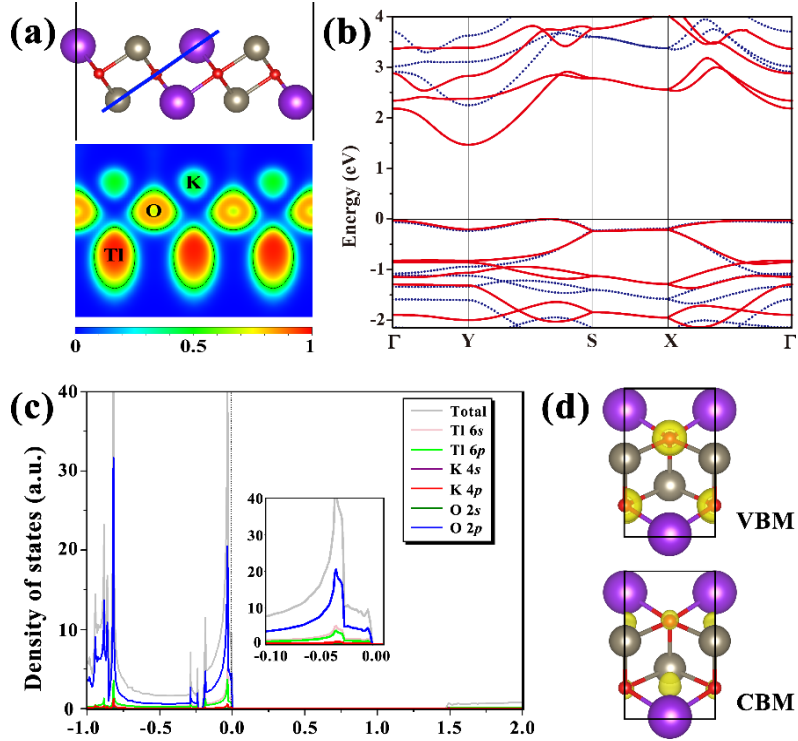

**Figure 3.** (a) Calculated electron localization function (ELF) of monolayer KTlO, where the blue line in the upper part denotes the location for the 2D cross section demonstration of the ELF (in the lower part). (b) Band structures of monolayer KTlO calculated using PBE (red solid line) and HSE06 (blue dotted line) functionals, without considering SOC. (c) Partial DOS of the KTlO monolayer calculated using the PBE functional. The inset highlights the DOS near the Fermi level. (d) Spatial distribution of the wave-functions corresponding to the VBM and CBM of monolayer KTlO.

As shown in **Fig. 3(b)**, monolayer KTlO exhibits an indirect band gap, calculated to be 1.47 eV and 2.25 eV according to PBE and HSE06 functionals. The CBM is located at the *Y*-point while the VBM lies along the *Y-M* direction, closer to the *M*-point. Note that a Mexican-hat-like dispersion was observed in monolayer KTlO, yielding a sharp one-dimensional-like van Hove singularity in the DOS around the Fermi level as shown in **Fig. 3(c)**. The partial DOS analysis shows that the states near the VBM mainly consist of O-3*p* orbitals, with a small contribution from Tl-6*s* and 6*p* orbitals. The spatial charge distributions of VBM and CBM are plotted in **Fig. 3(d)**. The VBM charges are mainly localized at the O atom, while the Tl atoms contribute most to the CBM states.

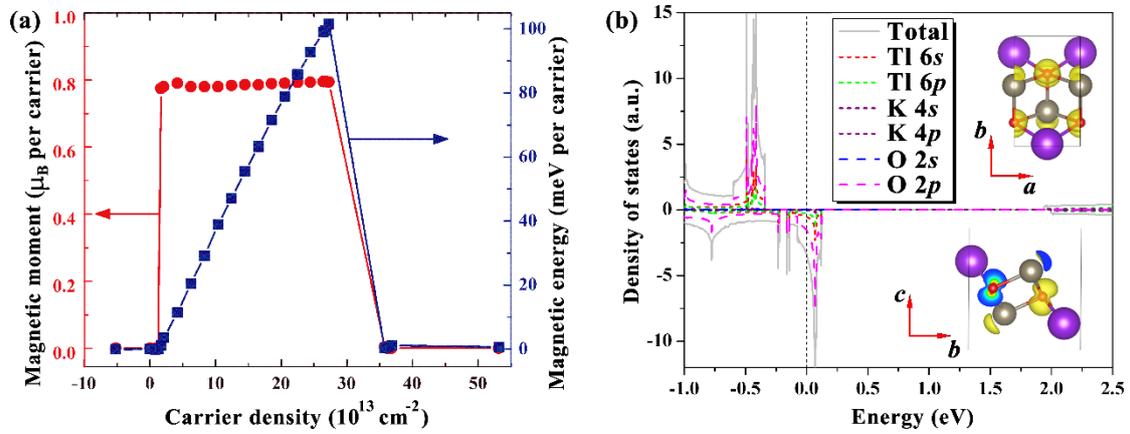

**Figure 4.** (a) Variation of the magnetic moment and magnetic energy (per carrier) of monolayer KTlO, upon different levels of carrier doping density. (b) Spin-resolved DOS of monolayer KTlO at a hole doping density of $2.73 \times 10^{14}$ cm$^{-2}$. The inset shows the spin density distribution, where apparently O atoms dominate the contribution.

Large DOS near the Fermi level may lead to electronic instabilities and possible phase transitions, such as magnetism and superconductivity.[34,34,35] In the case of monolayer KTlO, we find that electronic instability can be driven by the exchange interaction from hole doping. First, the 2D KTlO monolayer was doped with holes or electrons at various carrier concentrations. Our calculations show that it spontaneously develops a ferromagnetic ground state even for a tiny amount of hole doping. **Figure 4(a)** shows the calculated electron spin magnetic moment per carrier and the magnetic energy per carrier. Here the magnetic energy is defined as the total energy difference between the nonmagnetic and ferromagnetic states ($E_{Mag}=E_{NM}-E_{FM}$). Monolayer KTlO becomes ferromagnetic with a magnetic moment of 0.78 $\mu_B$/carrier at a very low carrier density 0.04 hole per cell ($1.67\times10^{13}$ cm$^{-2}$) considered in our calculations, i.e., with a minor spin-polarization energy (1.2 meV/carrier). Then the magnetic moment saturates at nearly 0.79 $\mu_B$/carrier upon increasing hole carrier density. The magnetic moment per carrier remains at this high value even though the spin carrier density increases up to a high carrier density of $2.73\times10^{14}$ cm$^{-2}$. Beyond this density, monolayer KTlO returns to a nonmagnetic state. Comparing to the magnetic moment which is nearly a constant above certain carrier density, the magnetic energy has a strong dependence on the doping level. Within GGA-PBE, the magnetic energy increases nearly monotonically to 101.6 meV/carrier at $2.73\times10^{14}$ cm$^{-2}$ doping level, followed by a sharp decrease back to zero at $3.57\times10^{14}$ cm$^{-2}$ doping level. Compared with monolayer GaSe (3 meV/carrier),[38] the highest magnetic energy of monolayer KTlO is ~30 times larger, suggesting its superiority for spintronics. The half-metal nature is clearly demonstrated in **Fig. 4(b)**, where the spin down states near the Fermi level are dominated by the 3*p* orbital of oxygen atoms, with a small contribution from Tl-6*s* and 6*p* orbitals. The corresponding spin density is primarily from the O and Tl atoms along the out-of-plane direction. It is even more interesting that a structural phase transition has been observed in the hole-doping process with a critical value of 0.68 hole per unit cell (*i.e.* $3.57\times10^{14}$ cm$^{-2}$), which is also the critical point for the ferromagnetic to nonmagnetic transition. The new phase is metastable and will collapse to the original structure when the doped holes disappear, indicating a repeatable hole-induced phase transition (details are shown in **Fig. S2**). On the other hand, monolayer KTlO does not induce any magnetism with electron doping, but results in a semiconductor-to-metal transition (see **Fig. S3**), which may be of interest for nanosensors.

In order to elucidate the variations in the electronic properties of KTlO from the bulk to few layers, we have investigated the electronic band gaps of 2D KTlO with varying number of layers. The results shown in **Fig. S4** are based on PBE calculations. The band gaps show a common trend of increase upon the attenuation of the film, with 1.20 eV, 1.24 eV, 1.28 eV and 1.31 eV for five-layer, four-layer, trilayer and bilayer KTlO, respectively. Similar to the bulk counterpart, all 2D KTlO multilayers under investigation have indirect band gaps. However, differences in detailed band diagram features can be observed. For bilayer KTlO, the CBM is located at the *Y*-point which is consistent with that of monolayer KTlO, while the VBM is located along the *Γ*-*X* direction. For tri-, four-, and five-layer KTlO, the band features are nearly the same, with the VBM located along *Γ*-*X* direction but the CBM deviated from *Y*. The possible reasons for band gap reduction and the band characteristics variation are the lacking of interlayer interaction as well as structural reconstruction upon stacking atomic layers.[57]

We further calculated the carrier mobilities (electrons and holes) of monolayer KTlO to explore its application potential in electron devices, based on the deformation potential theory proposed by Bardeen and Shockley.[58] The carrier mobility of 2D materials can be calculated by the following equation:[57,59,60]

$$\mu_{2D} = \frac{e\hbar^3 C_{2D}}{k_B T m^* m_d (E_l^i)^2},$$

where $\hbar$ is the reduced Planck constant, $k_B$ is the Boltzmann constant, $m^*$ is the effective mass in the transport direction, $m_d$ is the average effective mass determined by $m_d = (m_a^* m_b^*)^{1/2}$, and $T$ is the temperature (T = 300 K). The elastic modulus $C_{2D}$ of the longitudinal strain in the propagation direction is derived from $(E - E_0)/S_0 = C_{2D}(\Delta l / l_0)^2 / 2$, where $E$ is the total energy of the 2D structure, and $S_0$ is the lattice area of the equilibrium supercell. The deformation potential constant $E_1^i$ is defined as $E_1^i = \Delta E_i / (\Delta l / l_0)$. Here $\Delta E_i$ is the energy change of the $i^{th}$ band under proper cell compression and dilatation (calculated using a step of 0.5%), $l_0$ is the lattice constant in the transport direction and $\Delta l$ is the deformation of $l_0$.

**Table 1.** Calculated effective mass $m^*$ (unit: $m_e$), deformation potential constant $|E_1^i|$ (unit: eV), elastic modulus $C_{2D}$ (unit: N m$^{-1}$), carrier mobility $\mu_{2D}$ (unit: $10^3$ cm$^2$ V$^{-1}$s$^{-1}$) for monolayer KTlO along the $a$ (Y-M) direction and $b$ (Y-$\Gamma$) directions.

| Carrier type | $m_a^*$ | $m_b^*$ | $|E_{1a}|$ | $|E_{1b}|$ | $C_a^{2D}$ | $C_b^{2D}$ | $\mu_a^{2D}$ | $\mu_b^{2D}$ |
|---|---|---|---|---|---|---|---|---|
| Electron | 0.740 | 1.120 | 2.00 | 0.46 | 25.97 | 26.26 | 0.20 | 2.54 |
| Hole | 2.407 | 2.434 | 0.23 | 1.54 | 25.97 | 26.26 | 1.86 | 0.04 |

Our results are summarized in **Table 1**. The elastic moduli are 25.97 N m$^{-1}$ and 26.26 N m$^{-1}$ along the $a$ and $b$ directions, respectively. They are lower than those of Tl$_2$O ($C_a$ = 42.68 N m$^{-1}$ and $C_b$ = 28.57 N m$^{-1}$)[46]. The effective mass of an electron is anisotropic along the $a$ and $b$ directions (0.740 $m_e$ for $a$ and 1.120 $m_e$ for $b$). For the holes, the effective masses along $a$ (2.407 $m_e$) and $b$ direction (2.434 $m_e$) are nearly the same, and notably higher than the electrons. The large hole effective mass is due to the flat band near the VBM. The deformation potential constant $E_1$ shows obvious anisotropy both for the electron (1.90 and 0.26 eV) and the hole (0.46 eV and 1.54), similar to the case of Tl$_2$O.[46] The calculated electron mobility along the $b$ direction is 2.54×10$^3$ cm$^2$

V$^{-1}$ s$^{-1}$, which is about 13 times higher than that for the *a* direction (0.20×10$^3$ cm$^2$ V$^{-1}$s$^{-1}$), exhibiting strong anisotropy. Even stronger anisotropy is observed in the hole mobility, where the value along *a* (1.86×10$^3$ cm$^2$ V$^{-1}$ s$^{-1}$) is 46 times the value along *b* (0.04×10$^3$ cm$^2$ V$^{-1}$ s$^{-1}$), which suggests strongly direction-dependent conductivity.

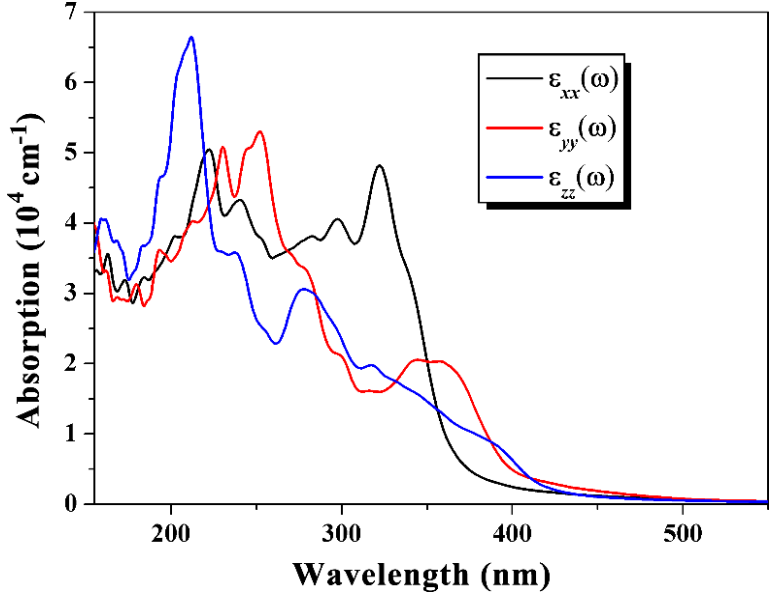

**Figure 5.** Calculated absorption coefficients of monolayer KTlO.

As the band gap value of monolayer KTlO fits the energy range of visible light, we further explored its optical properties by calculating the absorption spectra in- and out-of-plane using the HSE06 functional. The transverse dielectric function $\varepsilon(\omega) = \varepsilon_1(\omega) + i\varepsilon_2(\omega)$ is used to describe the optical properties of materials,[61] where $\omega$ is the photon frequency, $\varepsilon_1(\omega)$ is the real part and $\varepsilon_2(\omega)$ is the imaginary part of the dielectric function, respectively. The absorption coefficient can be evaluated according to the expression[61] $\alpha(\omega) = \frac{\sqrt{2}\omega}{c}\left\{\left[\varepsilon_1^2(\omega) + \varepsilon_2^2(\omega)\right]^{\frac{1}{2}} - \varepsilon_1(\omega)\right\}^{\frac{1}{2}}$. As shown in **Fig. 5**, the absorption coefficients of monolayer KTlO reach the order of 10$^4$ cm$^{-1}$, exhibiting optical absorption over a wide wave-length range in the visible and ultraviolet light region. In addition, in-plane optical absorption shows strong anisotropy along the *x* and *y* directions. The outstanding optical properties suggest potential applications of monolayer KTlO as efficient optical absorber materials in solar cells and optoelectronic devices.

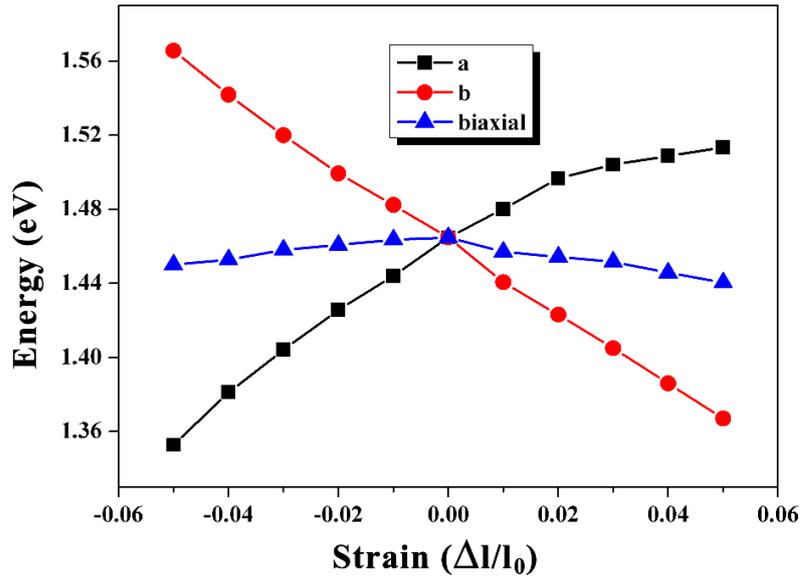

**Figure 6.** Electronic band gaps of monolayer KTlO under various applied strains, calculated using the PBE functional.

Finally, we studied the impact of in-plane compressive/tensile strains on the band structure of monolayer KTlO. **Figure 6** presents the PBE band gaps of monolayer KTlO under different strains in the range of -5% to 5% (the band structures are plotted in **Fig. S6**). Interestingly, the band gaps of monolayer KTlO decrease/increase monotonous with tensile/compressive strains, when it is subject to uniaxial strains along *a* or *b* axes. On the other hand, under the biaxial strain, the band gap value was insensitive to the magnitude of the strain, showing variations of less than 0.03 eV. In addition, the indirect bandgap feature remains unchanged within the range of strain being considered (see **Fig. S6**). The various responses to strain in monolayer KTlO may be of interest for future applications in flexible devices.

## IV. Conclusions

In summary, we propose that monolayer KTlO is a remarkable new 2D semiconductor for nanoelectronics and spintronic devices. The predicted cleavage energy of 0.56 J m$^{-2}$ indicates that exfoliation from the bulk is possible to produce monolayer KTlO. Furthermore, it possesses an indirect band gap of 2.25 eV with high carrier mobilities for electrons ($2.54 \times 10^3$ cm$^2$ V$^{-1}$ s$^{-1}$) and holes ($1.86 \times 10^3$ cm$^2$ V$^{-1}$ s$^{-1}$). In particular, we find that the 2D KTlO crystal shows electron instability in its band structure, and a non-magnetic to ferromagnetic transition can be achieved by moderate hole doping within a wide range of $1.67 \times 10^{13}$ cm$^{-2}$—$2.73 \times 10^{14}$ cm$^{-2}$. Such magnetic phase transition as well as its remarkable light absorption in the range of the visible and ultraviolet light region make it a promising candidate for spintronics and optoelectronic applications.

## Acknowledgement

This work was supported by the National Key Research and Development Program of China (Materials Genome Initiative, 2017YFB0701700), the National Natural Science Foundation of China under Grant No. 11704134, the Fundamental Research Funds of Wuhan City under Grant No. 2017010201010106, and the Fundamental Research Funds for the Central Universities of China under Grant No. HUST:2016YXMS212. K.-H. Xue received support from China Scholarship Council (No. 201806165012).

Supporting information for

# KTlO: A metal shrouded 2D semiconductor with high carrier mobility and tunable magnetism

Ya-Qian Song,[1,+] Jun-Hui Yuan,[1,+] Li-Heng Li,[1] Ming Xu,[1] Jia-Fu Wang,[2] Kan-Hao Xue,[1,3,*] and Xiang-Shui Miao[1]

[1] Wuhan National Research Center for Optoelectronics, School of Optical and Electronic Information, Huazhong University of Science and Technology, Wuhan 430074, China

[2] School of Science, Wuhan University of Technology, Wuhan 430070, China

[3] IMEP-LAHC, Grenoble INP – Minatec, 3 Parvis Louis Néel, 38016 Grenoble Cedex 1, France

[+]The authors Y.-Q. Song and J.-H. Yuan contributed equally to this work.

**Corresponding Author**

*E-mail: xkh@hust.edu.cn (K.-H. Xue)

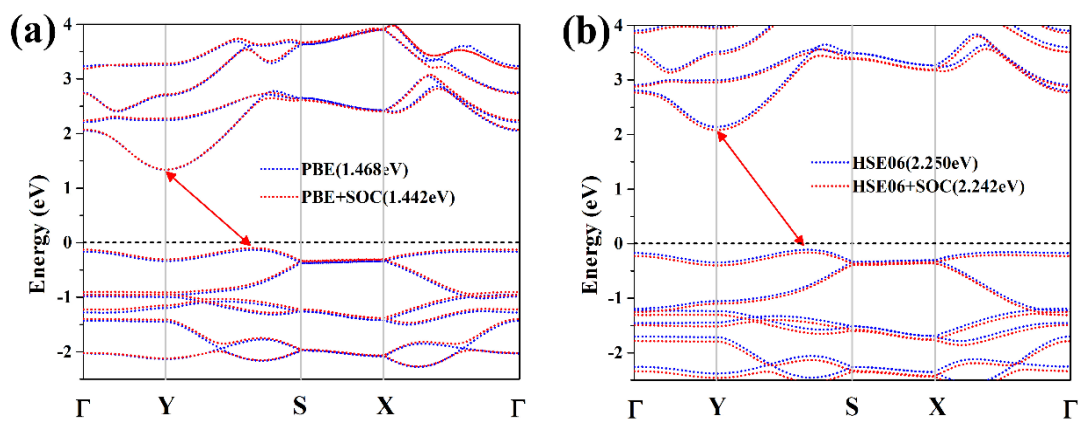

**Figure S1.** (a) Electronic band structures of monolayer KTlO at the PBE level, with or without considering spin-orbit coupling. (b) Electronic band structures of monolayer KTlO calculated using the screened HSE06 hybrid functional, with or without considering spin-orbit coupling.

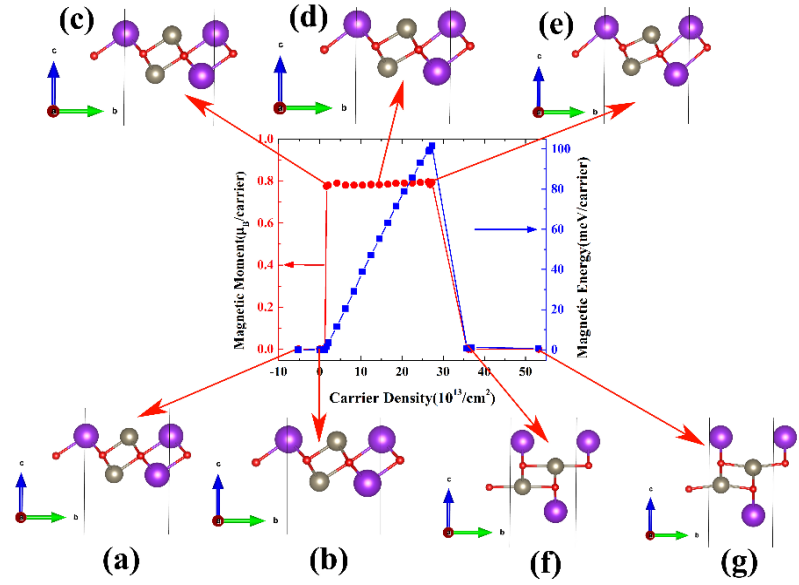

**Figure S2.** Crystal structures of monolayer KTlO versus various electron/hole-doping levels: (a) -5.2×10$^{13}$ cm$^{-2}$; (b) 0; (c) 1.67×10$^{13}$ cm$^{-2}$; (d) 16.54×10$^{13}$ cm$^{-2}$; (e) 27.3×10$^{13}$ cm$^{-2}$; (f) 35.7×10$^{13}$ cm$^{-2}$; (g) 53.2×10$^{13}$ cm$^{-2}$.

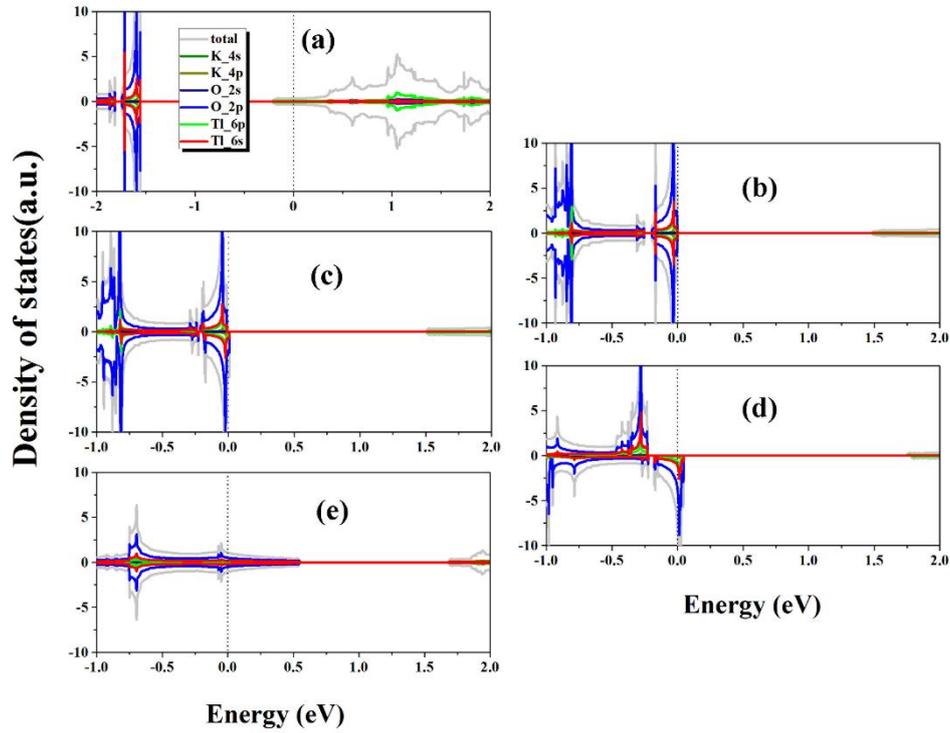

**Figure S3.** Projected density of states of monolayer KTlO versus various electron/hole-doping levels: (a) -5.2×10$^{13}$ cm$^{-2}$; (b) 0; (c) 1.67×10$^{13}$ cm$^{-2}$; (d) 16.54×10$^{13}$ cm$^{-2}$; (e) 36.83×10$^{13}$ cm$^{-2}$.

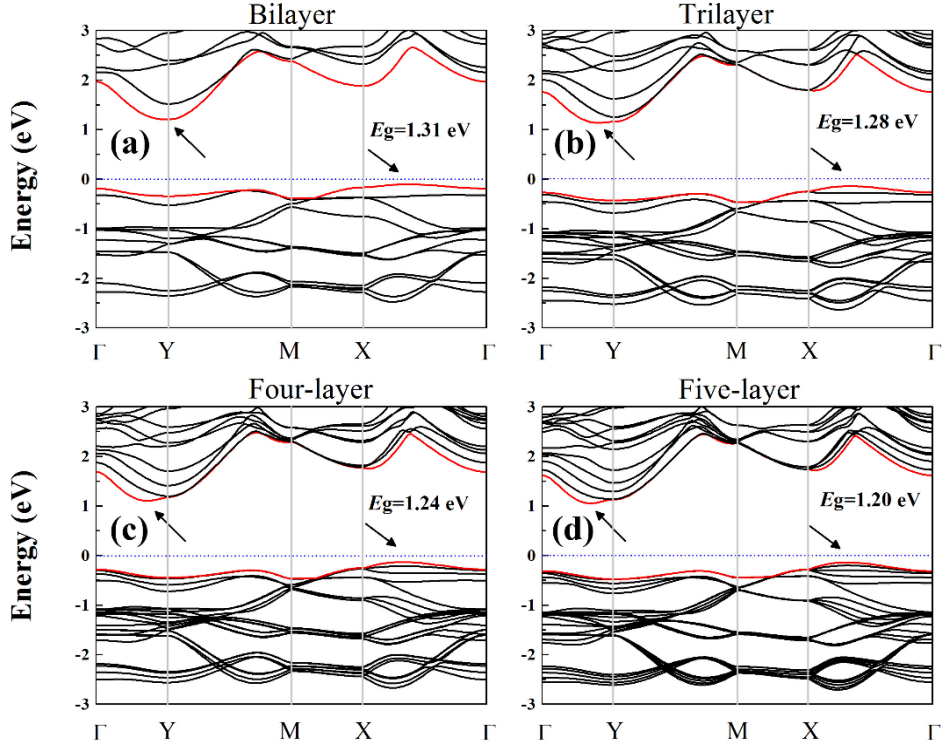

**Figure S4.** Electronic band structures of KTlO multilayers versus the number of atomic layers, calculated using the PBE functional: (a) bilayer; (b) trilayer; (c) four-layer; (d) five-layer.

**Table S1.** The carrier mobility $\mu_{2D}$ ($\times 10^3$ cm$^2$ V$^{-1}$s$^{-1}$) and energy bang gap (eV) at the HSE06 level (*d/i* represent direct/indirect band gaps) of KTlO, together with the values for Tl$_2$O and K$_2$O for comparison.

|  | Direction | KTlO | Tl$_2$O[1] | K$_2$O[2] |
|---|---|---|---|---|
| electron | *x* | 0.227 | 3.342 | 8.80 |
|  | *y* | 1.65 | 0.404 | 18.7 |
| hole | *x* | 1.37 | 4.302 | 0.005 |
|  | *y* | 0.04 | 0.016 | 0.0003 |
| band gap |  | 2.25 (*i*) | 1.56 (*d*) | ~2.4 (*i*) |

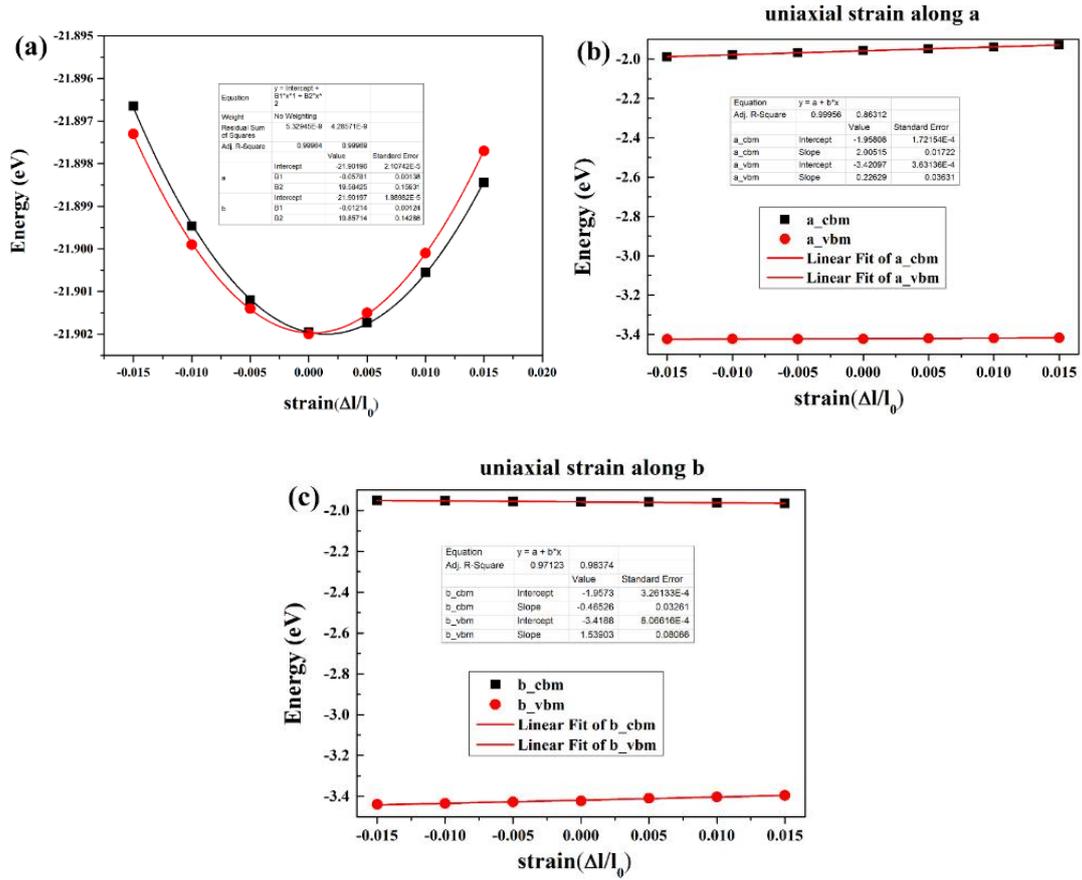

**Figure S5.** (a) Relation between total energy and the applied strain δ along the *a* (black curve) and *b* (red curve) directions of monolayer KTlO. The quadratic fitting of the data gives the in-plane stiffness of 2D structures. (b) The VBM/CBM shifts for monolayer KTlO with respect to the vacuum energy, as a function of the applied strain along the *a* direction. (c) The VBM/CBM shifts for monolayer KTlO with respect to the vacuum energy, as a function of the applied strain along the *b* direction. The linear fit of the data in (b) and (c) gives the deformation potential constants.

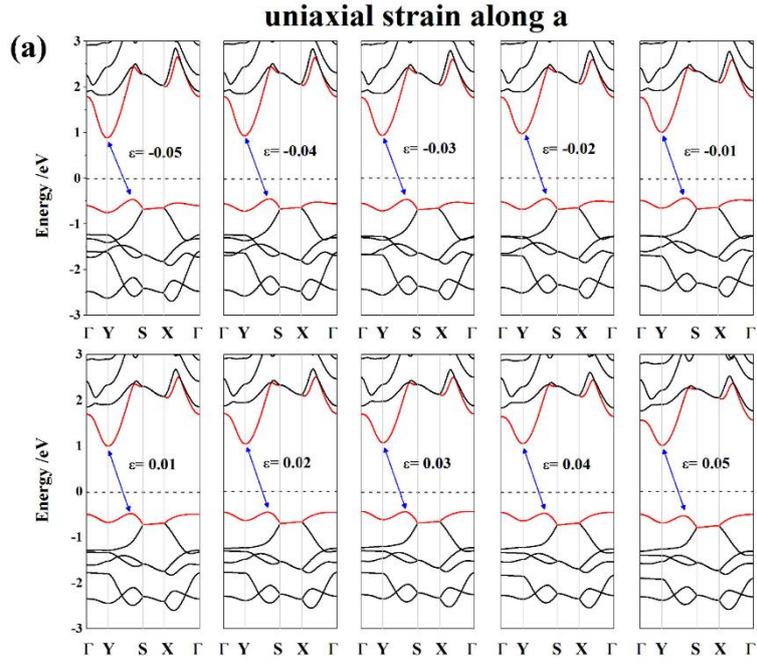

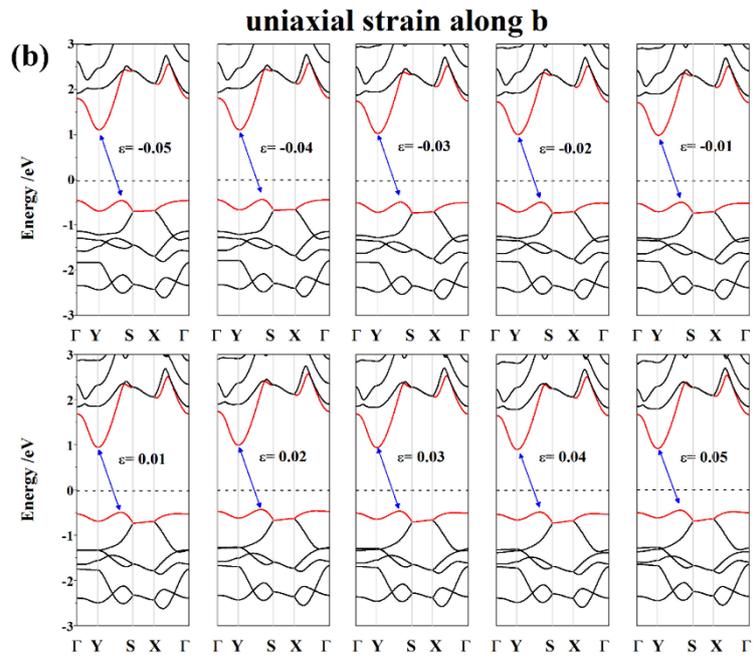

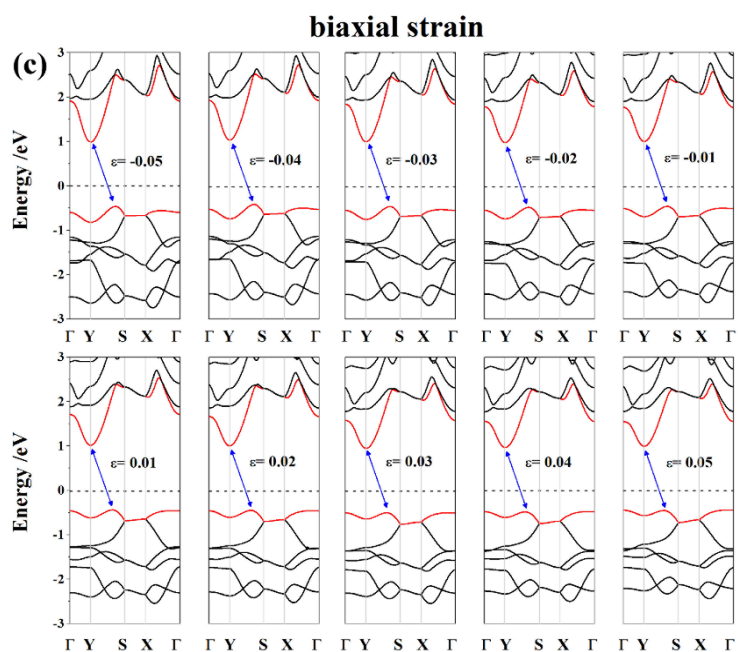

**Figure S6.** Electronic band structures of monolayer KTlO under various strain situations, calculated using the PBE functional. (a) Uniaxial strain along *a*-axis; (b) uniaxial strain along *b*-axis; (c) biaxial strain.

REFRENCES

(1) Ma, Y.; Kuc, A.; Heine, T. Single-Layer $Tl_2O$: A Metal-Shrouded 2D Semiconductor with High Electronic Mobility. *Journal of the American Chemical Society* **2017**, *139* (34), 11694–11697.
(2) Hua, C.; Sheng, F.; Hu, Q.; Xu, Z.-A.; Lu, Y.; Zheng, Y. Dialkali-Metal Monochalcogenide Semiconductors with High Mobility and Tunable Magnetism. **2018**.